\def\bfe{\mbox{\boldmath$\eta$}}
\def\bfz{\mbox{\boldmath$\zeta$}}
\def\bfx{\mbox{\boldmath$\xi$}}
\def\bfal{\mbox{\boldmath$\alpha$}}
\def\bfbt{\mbox{\boldmath$\beta$}}
\def\R{{\bf R}}
\def\C{{\bf C}}

\documentstyle[12pt,psfig]{article}
\begin{document}

\begin{center}
{\large\bf THE CENTER MANIFOLD THEOREM\\
FOR CENTER EIGENVALUES\\
WITH NON-ZERO REAL PARTS

}

\bigskip
O.M.Podvigina\footnote{E-mail: olgap@mitp.ru}

\bigskip
International Institute of Earthquake Prediction Theory\\
and Mathematical Geophysics,\\
79 bldg.~2, Warshavskoe ave., 117556 Moscow, Russian Federation

\bigskip
Laboratory of General Aerodynamics, Institute of Mechanics,\\
Lomonosov Moscow State University,\\
1, Michurinsky ave., 119899 Moscow, Russian Federation

\bigskip
Observatoire de la C\^ote d'Azur,\\
BP~4229, 06304 Nice Cedex 4, France

\end{center}

\begin{abstract}\noindent
We define center manifold as usual as an invariant manifold, tangent to the
invariant subspace of the linearization of the mapping defining a continuous
dynamical system, but the center subspace that we consider is associated with
eigenvalues with small {\it but not necessarily zero} real parts.
We prove existence and smoothness of such center manifold assuming that certain
inequalities between the center eigenvalues and the rest of the spectrum hold.
The theorem is valid for finite-dimensional systems,
as well as for infinite-dimensional systems provided they satisfy an additional
condition. We show that the condition holds
for the Navier-Stokes equation subject to appropriate boundary conditions.

\end{abstract}

\pagebreak
{\large\bf Introduction}

\bigskip
Investigation of bifurcations in complex dynamical systems, e.g.,
hydrodynamic or magnetohydrodynamic ones, can be simplified
by reducing dimension of the state space. This can be done by the
center manifold (CM) \cite{GucHol} or Lyapunov-Schmidt \cite{Gol85} reductions.
CM is an invariant manifold, tangent to an invariant subspace of
the linearization of the mapping defining the continuous dynamical system.
We will refer to the eigenvalues associated with the invariant subspace
as center eigenvalues. In conventional definitions of CM employed
in applications (e.g., \cite{arm,ChosIos,projo}) imaginary center eigenvalues
were assumed \cite{Hen,GucHol,Van89,VanI92}. Here we consider expanded CM,
allowing center eigenvalues with small {\it but not necessarily zero} real parts.

Our interest in such CM stems from the works \cite{PoAsHa,pod05}, where
they were applied for investigation of bifurcations in an ABC forced
hydrodynamic system. While the 6-dimensional reduced system, obtained by the
conventional CM reduction, reproduced only the first bifurcation of the trivial
steady state \cite{AshPod}, the 8-dimensional reduced system constructed with the use of an
expanded CM reproduced well the complex sequence of bifurcations of
the original hydrodynamic system \cite{PoAsHa,pod05}.

To the best of our knowledge, the variants of definitions of CM, where center
eigenvalues with real parts unequal to one\footnote{If a continuous system
is transformed into a discrete one by time discretization \cite{MarMc},
eigenvalues of linearization increase by 1, and thus in discrete dynamical
systems center eigenvalues have real parts close to 1.} were allowed, were
introduced before only for discrete finite-dimensional dynamical systems
\cite{Ioo,Shu}. Nontrivial
problems in the theory of CM are the questions of their existence and
smoothness. Theorems, guaranteeing existence and smoothness of CM for the
discrete finite-dimensional dynamical systems, where real parts of center
eigenvalues are close to one, are available \cite{Ioo,Shu}, but they
cannot be generalized by the standard technique \cite{MarMc} or other simple
arguments to cover the continuous infinite-dimensional case.

Our goal is to present a strict mathematical proof of the expanded CM
(for the sake of simplicity, we will henceforth refer to them without the
qualifier ``expanded'') theorem, which is applicable for hydrodynamic system.
First, the theorem is proved for finite-dimensional systems. Second,
we introduce a class of infinite-dimensional systems, for which the theorem
remains valid. Finally, we show that the Navier-Stokes equation belongs to
this class, if it is considered for appropriate boundary conditions and
provided certain inequalities hold for eigenvalues of the linearization of the
equation near the trivial steady state.

The theory which we develop here involves modifications of the proof of the CM
theorem for finite-dimensional systems \cite{Van89} (pp.~91-123), and of
generalization of this theorem for infinite-dimensional systems \cite{VanI92}
(pp.~126-160). We use a similar notation and follow the presentation of the
papers. If a theorem or a lemma proved in these papers is applied here in its
original form, we present only its statement. Our presentation is otherwise
complete.

\bigskip
{\large\bf 1. The center manifold theorem for center eigenvalues with
non-vanishing real parts. Finite-dimensional systems}

\bigskip
{\bf 1.1. The global CM theorem}

\bigskip
We consider differential equations of the form
\begin{equation}\label{equa}
\dot x=f(x)\equiv Ax+\tilde f(x),
\end{equation}
where $x\in\R^n$, $f:\R^n\to\R^n$ is a $C^k$ vector field, $k\ge1$, $f(0)=0$,
$A=Df(0)\in{\cal L}(\R^n)$ and hence $\tilde f(0)=0$, $D\tilde f(0)=0$.
For each $x\in\R^n$ we denote by $t\to\tilde x(t,x)$ the unique
solution to (\ref{equa}), satisfying $x(0)=x$; the maximal interval of its
existence is denoted by $J(x)$.
For an open $\Omega\subset\R^n$ and $x\in\Omega$ denote by $J_\Omega(x)$
the maximal interval of $t$ such that $\tilde x(\cdot,x)\in\Omega$.

Let the spectrum of the operator $A$,
$\sigma(A)\subset\C$, be decomposed as a disjoint union of the stable spectrum
$\sigma_s$, the center spectrum $\sigma_c$ and the unstable spectrum
$\sigma_u$, where
$$\sigma_s=\{\lambda\in\sigma\mid\hbox{Re}\lambda<-\Lambda^-\},$$
\begin{equation}\label{decomp}
\sigma_c=\{\lambda\in\sigma\mid-\Lambda^-\le\hbox{Re}\lambda\le\Lambda^+\},
\end{equation}
$$\sigma_u=\{\lambda\in\sigma\mid\hbox{Re}\lambda>\Lambda^+\}$$
and $\Lambda^\pm\ge0$.
Denote by $X_s$, $X_c$ and $X_u$ (the stable,
the center and the unstable subspaces) the subspaces of
$\R^n$ spanned by the generalized eigenvectors of $A$ associated with the
respective sets of eigenvalues; thus $\R^n=X_s\oplus X_c\oplus X_u$.
We call $X_h=X_s\oplus X_u$ the hyperbolic subspace. Denote by $\pi$
projections onto corresponding subspaces:
$$\pi_s:\R^n\to X_s,\quad\pi_c:\R^n\to X_c,\quad\pi_u:R^n\to X_u$$
and $\pi_h=\pi_s+\pi_u$.

Denote
\begin{equation}\label{albt}
\begin{array}{c}
\beta_+=\min\{\hbox{Re}\lambda\mid\lambda\in\sigma_u\}\\
\alpha_+=\max\{\hbox{Re}\lambda\mid\lambda\in\sigma_c\}\\
\alpha_-=-\min\{\hbox{Re}\lambda\mid\lambda\in\sigma_c\}\\
\beta_-=-\max\{\hbox{Re}\lambda\mid\lambda\in\sigma_s\}
\end{array}
\end{equation}
($\beta_+=+\infty$ if $\sigma_u=\emptyset$, and $\beta_-=+\infty$ if
$\sigma_s=\emptyset$). From (\ref{decomp}), $\beta_+>\alpha_+\ge0$ and
$\beta_->\alpha_-\ge0$.

\medskip
{\it Lemma 1.} For any $\epsilon>0$ there exists a constant $M(\epsilon)$
such that the following inequalities hold:
\begin{equation}
\begin{array}{ll}
\|e^{At}\pi_c\|\le M(\epsilon)e^{(\alpha_++\epsilon)t},&\quad\forall t\ge0,\\
\|e^{At}\pi_c\|\le M(\epsilon)e^{-(\alpha_-+\epsilon)t},&\quad\forall t\le0,\\
\|e^{At}\pi_u\|\le M(\epsilon)e^{(\beta_+-\epsilon)t},&\quad\forall t\le0,\\
\|e^{At}\pi_s\|\le M(\epsilon)e^{-(\beta_--\epsilon)t},&\quad\forall t\ge0.
\end{array}
\end{equation}
The proof is identical to the proof of Lemma 1.1 in \cite{Van89}
and it is omitted here.

\medskip
Denote by $C^k_b(X;Y)$ the set of all bounded mappings from a Banach space
$X$ to a Banach space $Y$ with the norm
$$\|w\|_{C^k_b}=\max_{0\le j\le k}|w|_j$$
where
$$|w|_j=\sup_{x\in X}\|D^jw(x)\|,$$
and $C^k_b(X;X)$ is denoted by $C^k_b(X)$.

Consider a system
\begin{equation}\label{equg}
\dot x=A x+g(x),
\end{equation}
where $x\in\R^n,\ A\in{\cal L}(\R^n)$ and $g\in C^k_b(\R^n)$ for some $k\ge1$.
Denote by $\tilde x_g(t,x)$ the solution to (\ref{equg}), satisfying $x(0)=x$.
Since $g$ is bounded, it is defined for all $t$.

\medskip
{\it Theorem 1.}
There exists $\delta_0>0$ (depending on $A\in{\cal L}(\R^n)$~) such
that for each $g\in C^1_b(R^n)$ with $|g|_1<\delta_0$ the following holds:

\medskip\noindent
(i) Existence and invariance: the set
\begin{equation}\label{cenman}
M_c=\{x\in\R^n\mid\sup_{t\in\R}\|\pi_h\tilde x_g(t,x)\|<\infty\}
\end{equation}
(which is called {\it global CM}) is invariant for (\ref{equg}).
It is also a $C^0$-submanifold in $\R^n$. More precisely, there exists
$\psi\in C^0_b(X_c;X_h)$ such that
\begin{equation}\label{manpsi}
M_c=\{x_c+\psi(x_c)\mid x_c\in X_c\};
\end{equation}

\noindent
(ii) Uniqueness: if $\phi\in C^0_b(X_c;X_h)$ is such that a manifold
$$W_\phi=\{x_c+\phi(x_c)\mid x_c\in X_c\}$$
is invariant under (\ref{equg}), then $W_\phi=M_c$ and $\phi=\psi$.

\medskip
The proof of invariance and uniqueness of $M_c$ is the same as in the proof
of Theorem 2.1 in \cite{Van89}, and we do not present it.
The proof of existence of $M_c$ follows.

\medskip
{\it Lemma 2.} Suppose $g\in C^1_b(R^n)$, $\eta_+\in(\alpha_+,\beta_+)$ and
$\eta_-\in(\alpha_-,\beta_-)$. Then
\begin{equation}\label{cmeta}
M_c=\{x\in\R^n\mid\max(\sup_{t>0}e^{-\eta_+t}\|\tilde x_g(t,x)\|,
\sup_{t<0}e^{\eta_-t}\|\tilde x_g(t,x)\|)<\infty\}.
\end{equation}

\medskip
{\it Proof.} The proof is based on the variation-of-constants formula
\begin{equation}\label{varcon}
\tilde x_g(t,x)=e^{A(t-t_0)}\tilde x_g(t_0,x)+
\int_{t_0}^t e^{A(t-\tau)}g(\tilde x_g(\tau,x))d\tau,
\end{equation}
which holds for all $t,t_0\in\R$.

First, we show that (\ref{cenman}) is a subset of (\ref{cmeta}).
Since $\eta_+>\alpha_+\ge0$ and $\eta_->\alpha_-\ge0$,
for $x$ from the set (\ref{cenman})
\begin{equation}\label{pih}
\sup_{t>0}e^{-\eta_+t}\|\pi_h\tilde x_g(t,x)\|<\infty,\ \hbox{and}\
\sup_{t<0}e^{\eta_-t}\|\pi_h\tilde x_g(t,x)\|)<\infty.
\end{equation}
Application of $\pi_c$ to (\ref{varcon}) with $t_0=0$ yields
\begin{equation}\label{pie}
\pi_c\tilde x_g(t,x)=e^{At}\pi_c x+
\int_0^t e^{A(t-\tau)}\pi_c g(\tilde x_g(\tau,x))d\tau.
\end{equation}
Lemma 1 implies that for $t>0$
$$\|\pi_c\tilde x_g(t,x)\|\le M(\eta_+-\alpha_+)e^{\eta_+t}\|x\|+
M(\eta_+-\alpha_+)\|g\|_0\int_0^t e^{\eta_+(t-\tau)}d\tau$$
$$\le M(\eta_+-\alpha_+)e^{\eta_+t}(\|x\|+\eta_+^{-1}\|g\|_0)$$
and hence
\begin{equation}\label{pie+}
\sup_{t>0}e^{-\eta_+t}\|\pi_c\tilde x_g(t,x)\|<\infty.
\end{equation}
It can be shown similarly that
$$\sup_{t<0}e^{\eta_-t}\|\pi_c\tilde x_g(t,x)\|<\infty,$$
which together with (\ref{pih}) and (\ref{pie+}) yields
$$\max(\sup_{t>0}e^{-\eta_+t}\|\tilde x_g(t,x)\|,
\sup_{t<0}e^{\eta_-t}\|\tilde x_g(t,x)\|)<\infty.$$

Conversely, assume that $x\in\R^n$ is from the set (\ref{cmeta}).
Project (\ref{varcon}) onto $X_u$ to obtain
\begin{equation}\label{proju}
\pi_u\tilde x_g(t,x)=e^{A(t-t_0)}\pi_u\tilde x_g(t_0,x)+
\int_{t_0}^t e^{A(t-\tau)}\pi_u g(\tilde x_g(\tau,x))d\tau.
\end{equation}
For a fixed $t\in\R$, $t_0\ge\max(0,t)$ and $\epsilon\in(0,\beta_+-\eta_+)$
Lemma 1 and (\ref{cmeta}) imply
$$\|e^{A(t-t_0)}\pi_u\tilde x_g(t_0,x)\|\le M(\epsilon)
e^{(\beta_+-\epsilon)(t-t_0)}Ce^{\eta_+t_0}$$
\begin{equation}\label{pu1}
=M(\epsilon)Ce^{(\beta_+-\epsilon)t}e^{-(\beta_+-\eta_+-\epsilon)t_0}.
\end{equation}
The r.h.s. of (\ref{pu1}) tends to zero when $t_0\to\infty$. Consequently,
in the limit $t_0\to\infty$ (\ref{proju}) takes the form
\begin{equation}\label{pinfty}
\pi_u\tilde x_g(t,x)=-\int_t^{\infty}
e^{A(t-\tau)}\pi_u g(\tilde x_g(\tau,x))d\tau,\quad\forall t\in\R.
\end{equation}
Thus, for any $\epsilon\in(0,\beta_+)$ and any $t\in\R$
\begin{equation}\label{puxg}
\|\pi_u\tilde x_g(t,x)\|\le M(\epsilon)\|g\|_0\int_t^{\infty}
e^{(\beta_+-\epsilon)(t-\tau)}d\tau=(\beta_+-\epsilon)^{-1}M(\epsilon)\|g\|_0.
\end{equation}
Similarly, for any $\epsilon\in(0,\beta_-)$ and any $t\in\R$
\begin{equation}\label{psinfty}
\pi_s\tilde x_g(t,x)=\int_{-\infty}^t
e^{A(t-\tau)}\pi_s g(\tilde x_g(\tau,x))d\tau
\end{equation}
and
\begin{equation}\label{psxg}
\|\pi_s\tilde x_g(t,x)\|\le(\beta_--\epsilon)^{-1}M(\epsilon)\|g\|_0.
\end{equation}
Together, (\ref{puxg}) and (\ref{psxg}) imply (\ref{cenman}).
The proof of Lemma 2 is completed.

\medskip
{\it Definition 1.} For a vector $\bfe=(\eta_+,\eta_-)$,
where $\eta_+,\eta_-\ge0$, $Y_{\bfe}$ is the Banach space
\begin{equation}\label{yeta}
Y_{\bfe}=\{y\in C^0(\R;\R^n)\mid\|y\|_{\bfe}=
\sup_{t\in\R}e^{-\bfe(t)}\|y(t)\|<\infty\},
\end{equation}
where
\begin{equation}\label{defet}
\bfe(t)=\left\{\begin{array}{ll}
\eta_+t&\quad\hbox{if}\ \ t\ge0,\\
-\eta_-t&\quad\hbox{if}\ \ t<0.
\end{array}
\right.
\end{equation}
The inequality $\bfz\ge\bfe$ means that $\zeta_+\ge\eta_+$ and
$\zeta_-\ge\eta_-$, and $\bfz>\bfe$ -- that $\zeta_+>\eta_+$ and
$\zeta_->\eta_-$. $Y_{\bfe}$ are a scale of Banach spaces: if $\bfz\ge\bfe$,
then $Y_{\bfe}\subset Y_{\bfz}$, and the embedding is continuous
$$\|y\|_{\bfz}\le\|y\|_{\bfe},\quad\forall y\in Y_{\bfe}.$$

In this notation, the manifold (\ref{cmeta}) can be expressed as
$$M_c=\{ x\in\R^n\mid\tilde x_g(\cdot,x)\in Y_{\bfe}\}$$
\begin{equation}\label{mcy0}
=\{ y(0)\mid y\in Y_{\bfe}\quad\hbox{and}\ y\ \hbox{solves}\ (\ref{equg})\}
\end{equation}
for some
$$\bfe\in(\alpha_+,\beta_+)\times(\alpha_-,\beta_-).$$

The scale of Banach spaces $Y_{\eta}$, $\eta>0$, employed in the proof of
the conventional CM theorem \cite{Van89}, coincides with the scale
(\ref{yeta}), where $\eta=\eta_-=\eta_+$; the spaces for $0<\eta<\beta$
are employed, where $\beta=\min(\beta_+,\beta_-)$ (cf. (\ref{yeta})
for $\alpha_+=\alpha_-=0$).

As it was shown in the proof of Lemma 2, (\ref{pie}), (\ref{pinfty}) and
(\ref{psinfty}) hold for $\tilde x_g(t,x)$ on the CM. Summing up these
equations we find that $x\in\R^n$ belongs to $M_c$ if and only if
$\forall t\in\R$
$$\tilde x_g(t,x)=e^{At}\pi_c x+
\int_0^t e^{A(t-\tau)}\pi_c g(\tilde x_g(\tau,x))d\tau+$$
$$\int_{-\infty}^{+\infty}B(t-\tau)\ g(\tilde x_g(\tau,x))d\tau,$$
where $B:\R\to{\cal L}(\R^n)$ is
\begin{equation}
B(t)=\left\{\begin{array}{ll}
-e^{At}\pi_u,&\quad\hbox{if}\ t<0,\\
e^{At}\pi_s,&\quad\hbox{if}\ t\ge0.
\end{array}
\right.
\end{equation}
Lemma 1 implies that for any $\epsilon>0$
\begin{equation}\label{normab}
\|B(t)\|<\left\{\begin{array}{ll}
M(\epsilon)e^{(\beta_+-\epsilon)t},&\quad\forall t<0,\\
M(\epsilon)e^{-(\beta_--\epsilon)t},&\quad\forall t>0.
\end{array}
\right.
\end{equation}

\medskip
{\it Lemma 3.} Suppose $g\in C^1_b(R^n)$,
$\bfe\in(\alpha_+,\beta_+)\times(\alpha_-,\beta_-)$ and $y\in Y_{\bfe}$. Then
$y$ is a solution to (\ref{equg}) if and only if there exists $x_c\in X_c$,
such that for any $t\in\R$
\begin{equation}\label{exix}
y(t)=e^{At}x_c+\int_0^t e^{A(t-\tau)}\pi_c g(y(\tau))d\tau+
\int_{-\infty}^{+\infty}B(t-\tau)\ g(y(\tau))d\tau.
\end{equation}
The proof is identical to the proof of Lemma 2.8 in \cite{Van89}
and it is omitted here.

\medskip
Let $\Sigma$ be the set of all $(x_c,y)\in X_c\times Y_{\bfe}$ such
that (\ref{exix}) holds; (\ref{mcy0}) implies
\begin{equation}\label{msigma}
M_c=\{y(0)\mid(x_c,y)\in\Sigma\}=\{x_c+\pi_h y(0)\mid(x_c,y)\in\Sigma\},
\end{equation}
since $\pi_c y(0)=x_c$ for any $(x_c,y)\in\Sigma$. To determine the set
$\Sigma$, rewrite (\ref{exix}) in the form
\begin{equation}\label{equy}
y=Sx_c+KG(y)
\end{equation}
where the following notation is used:
$$Sx_c:\R\to\R^n,\qquad (Sx_c)(t)=e^{At}x_c\quad\forall x_c\in X_c;$$
$$G(y):\R\to\R^n,\qquad G(y)(t)=g(y(t))\quad\hbox{for each function }y:\R\to\R^n;$$
\begin{equation}\label{defk}
Ky:\R\to\R^n,\quad Ky(t)=\int_0^t e^{A(t-\tau)}\pi_c y(\tau)d\tau+
\int_{-\infty}^{+\infty}B(t-\tau)\ y(\tau)d\tau
\end{equation}
for such functions $y:\R\to\R^n$ that the integrals are defined.

\medskip
{\it Lemma 4.} $S$ is a bounded operator from $X_c$ to $Y_{\bfe}$ for
any $\eta_+>\alpha_+$ and $\eta_->\alpha_-$.

\medskip
{\it Proof}. Lemma 1 implies that for any $\eta_+>\alpha_+$
$$\|e^{At}x_c\|\le M(\eta_+-\alpha_+)e^{\eta_+t}\|x_c\|,\quad\forall t>0,$$
and for any $\eta_->\alpha_-$
$$\|e^{At}x_c\|\le M(\eta_--\alpha_-)e^{-\eta_-t}\|x_c\|,\quad\forall t<0.$$
Hence
$$\|Sx_c\|_{\bfe}\le\max(M(\eta_+-\alpha_+),M(\eta_--\alpha_-))
\|x_c\|,\quad\forall x_c\in X_c.$$

\medskip
{\it Lemma 5.} If $g\in C^0_b(\R^n)$, then $G$ maps $C^0(\R;\R^n)$ into
$C^0_b(\R;\R^n)$, and $G$ maps each $Y_{\bfe}$, $\bfe\ge 0$, into itself.
If $g\in C^1_b(\R^n)$, then for any $\bfe>0$
$$\|G(y_1)-G(y_2)\|_{\bfe}\le|g|_1\|y_1-y_2\|_{\bfe},
\quad\forall y_1,y_2\in Y_{\bfe}.$$
\medskip
{\it Proof}. The first part is obvious. If $g\in C^1_b(\R^n)$,
$y_1,y_2\in Y_{\bfe}$, then
$$\sup_{t>0}e^{-\eta_+t}\|G(y_1)-G(y_2)\|=
\sup_{t>0}e^{-\eta_+t}\|g(y_1(t))-g(y_2(t))\|$$
$$\le\sup_{t>0}e^{-\eta_+t}|g|_1\|y_1(t)-y_2(t)\|\le
|g|_1\|y_1(t)-y_2(t)\|_{\bfe}.$$
A similar inequality holds for negative $t$. Thus, by virtue of (\ref{yeta})
and (\ref{defet}), the proof is complete.

\medskip
{\it Lemma 6.} For any $\bfe\in(\alpha_+,\beta_+)\times(\alpha_-,\beta_-)$
the operator $K:Y_{\bfe}\to Y_{\bfe}$ is bounded; there exists a continuous
function $\gamma:(\alpha_+,\beta_+)\times(\alpha_-,\beta_-)\to\R$ such that
\begin{equation}\label{lem6}
\|K\|_{\bfe}\le\gamma(\bfe),\quad\forall
\bfe\in(\alpha_+,\beta_+)\times(\alpha_-,\beta_-).
\end{equation}
\medskip
{\it Proof}. Suppose $\eta_+\in(\alpha_+,\beta_+)$, $\eta_-\in(\alpha_-,\beta_-)$,
$y\in Y_{\bfe}$ and $t>0$. The definition of $K$
(\ref{defk}) and bounds (\ref{normab}) imply
$$e^{-\eta_+t}\|Ky(t)\|\le\|y\|_{\bfe}\sup_{t>0}e^{-\eta_+t}\left[
\int_0^t\|e^{A(t-\tau)}\pi_c\|e^{\eta_+\tau}d\tau+
\int_{-\infty}^0\|B(t-\tau)\|e^{-\eta_-\tau}d\tau\right.$$
$$+\left.\int_0^t\|B(t-\tau)\|e^{\eta_+\tau}d\tau+
\int_t^{+\infty}\|B(t-\tau)\|e^{\eta_+\tau}d\tau\right]$$
$$\le\|y\|_{\bfe}\sup_{t>0}\left[
\int_0^t\|e^{A(t-\tau)}\pi_c\|e^{-\eta_+(t-\tau)}d\tau+
e^{(-\eta_+-\eta_-)t}\int_{-\infty}^0
\|B(t-\tau)\|e^{\eta_-(t-\tau)}d\tau\right.$$
$$+\left.\int_0^t\|B(t-\tau)\|e^{-\eta_+(t-\tau)}d\tau+
\int_t^{+\infty}\|B(t-\tau)\|e^{-\eta_+(t-\tau)}d\tau\right]$$
$$\le\|y\|_{\bfe}\left[
\int_0^{+\infty}\|e^{A\tau}\pi_c\|e^{-\eta_+\tau}d\tau+
\int_0^{+\infty}\|B(\tau)\|e^{\eta_-\tau}d\tau\right.$$
\begin{equation}\label{plus}
\left.+\int_0^{+\infty}\|B(\tau)\|e^{-\eta_+\tau}d\tau+
\int_{-\infty}^0\|B(\tau)\|e^{-\eta_+\tau}d\tau\right]
\end{equation}
$$\le\|y\|_{\bfe}\left[M(\epsilon_1)(\eta_+-\alpha_+-\epsilon_1)^{-1}+
M(\epsilon_2)(\beta_--\eta_--\epsilon_2)^{-1}\right.$$
$$+\left.M(\epsilon_3)(\beta_-+\eta_+-\epsilon_3)^{-1}+
M(\epsilon_4)(\beta_+-\eta_+-\epsilon_4)^{-1}\right],$$
if $\epsilon_i$ satisfy
$\eta_+-\alpha_+-\epsilon_1>0$, $\beta_--\eta_--\epsilon_2>0$,
$\beta_-+\eta_+-\epsilon_3>0$ and $\beta_+-\eta_+-\epsilon_4>0$.
Similarly, for $t<0$
\begin{equation}\label{minus}
e^{\eta_-t}\|Ky(t)\|\le\|y\|_{\bfe}\left[
\int_{-\infty}^0\|e^{A\tau}\pi_c\|e^{\eta_-\tau}d\tau
+\int_{-\infty}^0\|B(\tau)\|e^{-\eta_+\tau}d\tau\right.
\end{equation}
$$\left.+\int_{-\infty}^0\|B(\tau)\|e^{\eta_-\tau}d\tau+
\int_0^{+\infty}\|B(\tau)\|e^{\eta_-\tau}d\tau\right].$$
Thus $K\in{\cal L}(Y_{\bfe})$. The norm of $K$ is bounded by the function
$\gamma(\eta_+,\eta_-)$, defined as the maximum of the sums (\ref{plus}) and
(\ref{minus}); this is a continuous function of the two arguments.
The proof of Lemma 6 is complete.

\medskip
{\it Lemma 7.} If $\bfe\in(\alpha_+,\beta_+)\times(\alpha_-,\beta_-)$ and
$g\in C^1_b(\R^n)$ is such that
\begin{equation}\label{lem7}
\kappa=\|K\|_{\bfe}|g|_1<1
\end{equation}
then $(I-K\circ G)$ is a homeomorphism on $Y_{\bfe}$, whose inverse
$\Psi:Y_{\bfe}\to Y_{\bfe}$ is Lipschitzian with the Lipschitz constant
$\kappa$, and
\begin{equation}\label{solut}
\Sigma=\{(x_c,\Psi(Sx_c))\mid x_c\in X_c\}.
\end{equation}
The proof is identical to the proof of Lemma 2.12 in \cite{Van89}.

\medskip
We finish now the proof of Theorem 1. For a $\gamma(\bfe)$ satisfying
(\ref{lem6}), denote
$$\delta_0=\sup_{\bfe\in (\alpha_+,\beta_+)\times(\alpha_-,\beta_-)}
\gamma(\bfe)^{-1}.$$
If $g\in C^1_b(\R^n)$ and $|g|_1<\delta_0$, there exists
$\bfe\in(\alpha_+,\beta_+)\times(\alpha_-,\beta_-)$ such that
$|g|_1\gamma(\bfe)<1$. By (\ref{lem6}) this implies (\ref{lem7})
and therefore (\ref{solut}) holds. Combining it with (\ref{msigma}), obtain
(\ref{manpsi}) with $\psi:X_c\to X_h$ defined by
\begin{equation}\label{manend}
\psi(x_c)=\pi_h\Psi(Sx_c)(0),\quad\forall x_c\in X_c.
\end{equation}
Since $\Psi$ is continuous, $\psi$ is also continuous. Moreover, since
$\Psi=(I-K\circ G)^{-1}$ by definition,
$$\Psi(Sx_c)=Sx_c+KG(\Psi(Sx_c)).$$
From the definitions of $S$, $G$ and $K$ it follows that
$$\psi(x_c)=\int_{-\infty}^{+\infty}B(-\tau)g(Sx_c)(\tau)d\tau,$$
Thus, the bounds (\ref{normab}) imply
$$\|\psi(x_c)\|<(M(\epsilon_+)(\beta_+-\epsilon_+)^{-1}+
M(\epsilon_-)(\beta_--\epsilon_-)^{-1})|g|_0$$
$$\forall x_c\in X_c,\ \forall\epsilon_+\in(\alpha_+,\beta_+),\
\forall\epsilon_-\in(\alpha_-,\beta_-).$$
Finally, note that $\psi\in C^0_b(\R^n)$ is globally Lipschitzian, because
$\Psi$ is globally Lipschitzian and (\ref{manend}) holds. The Theorem is proved.

\bigskip
{\bf 1.2. Smoothness of CM}

\bigskip
The Theorem can be applied to study bifurcations, if a CM
is sufficiently smooth. In the sequel we prove
smoothness of the manifold under certain additional assumptions.

\medskip
{\it Theorem 2.} Let the spectrum of $A\in{\cal L}(\R^n)$ in (\ref{equg})
be split as
$\sigma(A)=\sigma_u\cup\sigma_c\cup\sigma_s$ in accordance with (\ref{decomp}),
with $\alpha_\pm$ and $\beta_\pm$ (see (\ref{albt})~) satisfying
$$\alpha_+<\beta_+/l\quad\hbox{and}\quad\alpha_-<\beta_-/l$$
for some $l\ge1$. Then for each $k$, $1\le k\le l$, there exists
$\delta_k\in(0,\delta_0]$ such that if $g\in C^k_b(\R^n)$ and $|g|_1<\delta_k$,
then the unique global center manifold $M_c$ of (\ref{equg}) is
$C^k$. More precisely, the mapping $\psi$ constructed in Theorem 1
belongs to $C^k_b(X_c,X_h)$.

\medskip
Since $\psi(x_c)=\pi_h\Psi(Sx_c)(0)$, it is sufficient to show that the
mapping $\Psi$ constructed in Lemma 7 is $C^k$. Then the smoothness might be
established by application of the implicit function theorem
to the equation (\ref{equy}), if the operator $G$ were $C^k$. The
difficulty is that as a mapping from
$Y_{\bfe}$ into itself $G$ is not in general differentiable. But
$G\in C^k(Y_{\bfe},Y_{\bfz})$, if $g\in C^k_b(\R^n)$ and $\bfz>k\bfe$.
The proof of Theorem 2 employs the following Lemma.

\medskip
{\it Lemma 8.} Suppose $g\in C^k_b(\R^n)$ for some $k\ge1$. Let
$\bfe,\bfz\in(\alpha_+,\beta_+)\times(\alpha_-,\beta_-)$ be such that
$\bfz>k\bfe$. Suppose
\begin{equation}\label{lem8}
\kappa=\sup_{\bfx\in[{\bfe},{\bfz}]}\|K\|_{\bfx}|g|_1<1.
\end{equation}
Then the mapping $\Psi:Y_{\bfe}\to Y_{\bfe}$ constructed in Lemma 7 belongs to
$C^k(Y_{\bfe},Y_{\bfz})$. More precisely,
$$\Psi-J_{\bfe,\bfz}\in C^k_b(Y_{\bfe};Y_{\bfz}),$$
where $J_{\bfe,\bfz}$ is the embedding of $Y_{\bfe}$ into $Y_{\bfz}$.

The proof of this Lemma coincides with the proof of Lemma 3.2 of \cite{Van89}
(pp.~104-115) after replacement of $\eta$, $\zeta$ and $\xi$ by
$\bfe$, $\bfz$ and $\bfx$, respectively. We do not repeat it here.

\medskip
{\it Proof of Theorem 2.} For each $k\ge 1$ denote
$$\delta_k=\sup_{\bfe\in(\bfal,\bfbt/k)}\ \inf_{\bfx\in[\bfe,k\bfe]}\gamma(\bfx)^{-1},$$
where $\gamma$ is the function constructed in Lemma 6. If $g\in C^k_b(\R^n)$
and $|g|_1<\delta_k$, then there exists $\bfe\in(\bfal,\bfbt/k)$ such
that $|g|_1<\inf\{\gamma(\bfx)^{-1}\mid\bfx\in[\bfe,k\bfe ]\}$. Since $\gamma$
is continuous, this implies existence of $\bfz\in(k\bfe,\bfbt)$ such that
$|g|_1<\inf\{\gamma(\bfx)^{-1}\mid\bfx\in [\bfe,\bfz ]\}$. By (\ref{lem6})
this implies (\ref{lem8}). It follows from Lemma 8 that $\Psi\in
C^k(Y_{\bfe},Y_{\bfz})$, all its derivatives being
globally bounded. Since $S:X_c\to Y_{\bfe}$ is a bounded linear operator
(Lemma 4), the mapping $x_c\to\Psi(Sx_c)$ is also $C^k(X_c,Y_{\bfz})$
with all its derivatives globally bounded. Hence (\ref{manend})
implies that $\psi\in C^k_b(X_c;X_h)$. The proof is complete.

\bigskip
{\bf 1.3. The local CM theorem}

\bigskip
Theorems 1 and 2 hold for all functions $g$, bounded by certain constants.
Now let us return to the equation (\ref{equa}), where $f$ does not
satisfy this condition.

\medskip
{\it Theorem 3.} Suppose $f\in C^k(\R^n)$, $k\ge1$, and $f(0)=0$.
Split the set of eigenvalues of $A=Df(0)$ in agreement with (\ref{decomp}):
$\sigma(A)=\sigma_u\cup\sigma_c\cup\sigma_s$. Assume
\begin{equation}\label{abk}
\alpha_+<\beta_+/k\quad\hbox{and}\quad\alpha_-<\beta_-/k
\end{equation}
($\alpha_\pm$ and $\beta_\pm$ are defined by (\ref{albt})~). Then there exists
$\psi\in C^k_b(X_c,X_h)$ ($X_c$ and $X_h$ denote the respective
center and hyperbolic subspaces) and an open neighborhood $\Omega$ of
the origin in $\R^n$ such that

\medskip\noindent
(i) $\psi(0)=0$ and $D\psi(0)=0$;

\noindent
(ii) the manifold
$$W_\psi=\{x_c+\psi(x_c)\mid x_c\in X_c\}$$
is locally invariant for (\ref{equa}), i.e.
$$\tilde x(t,x)\in W_\psi,\quad\forall x\in W_\psi\cap\Omega,\ \forall t\in
J_\Omega(x)$$

\noindent
(iii) if $x\in\Omega$ and $J_\Omega(x)=\R$, then $x\in W_\psi$.

\medskip
To prove Theorem 3, apply Theorems 1 and 2 to the system
\begin{equation}\label{equft}
\dot x=Ax+\tilde f_\rho(x),
\end{equation}
where
$$\tilde f_\rho(x)=\tilde f(x)\chi(\rho^{-1}x),\quad\forall x\in\R^n,$$
and $\chi$ is a smooth cut-off function $\chi:\R^n\to\R$ with the following
properties:

\noindent
(i) $0\le\chi\le 1,\ \forall x\in\R^n;$

\noindent
(ii) $\chi(x)=1,$ if $\|x\|\le 1;$

\noindent
(iii) $\chi(x)=0,$ if $\|x\|\ge 2.$

\noindent
The constant $\rho$ can be chosen small enough, so that the system (\ref{equft})
satisfies conditions of the Theorems. Equations (\ref{equa}) and
(\ref{equft}) coincide in $\Omega=\{x\in\R^n\mid\|x\|\le\rho\}$.

If $x\in\Omega$ and $J_\Omega(x)=\R$, then
$\tilde x(\cdot,x)=\tilde x_\rho(\cdot,x)$ is a bounded solution to
(\ref{equft}) and, according to (\ref{cenman}), belongs to its global CM,
thus implying $x\in W_\psi$.

\medskip
If the unstable spectrum is empty, the CM is attracting.
The proof is the same as in the case of a conventional CM theorem.

\bigskip
{\large\bf 2. CM theorem for infinite-dimensional systems}

\bigskip
Let $X$, $Y$ and $Z$ be Banach spaces with $X$ continuously embedded in $Y$, and
$Y$ continuously embedded in $Z$. Consider a differential equation
\begin{equation}\label{equa2}
\dot x=Ax+g(x),
\end{equation}
where $A\in{\cal L}(X,Z)$ and $g\in C^k(X,Y)$, $k\ge1$.

\medskip
{\it Definition 2.} For a vector $\bfe=(\eta_+,\eta_-)$, where $\eta_\pm\ge0$,
and a Banach space $E$, define a Banach space $BC^{\bfe}(\R,E)$:
\begin{equation}\label{yeta2}
BC^{\bfe}(\R,E)=\{w\in C^0(\R;E)\mid\|w\|_{\bfe}=
\sup_{t\in\R}e^{-\bfe(t)}\|w(t)\|_E<\infty\},
\end{equation}
where $\bfe(t)$ is defined by (\ref{defet}).

\bigskip
Assume the operator $A$ satisfies the following hypothesis ($\bf H$):

\bigskip
There exists a continuous projection $\pi_c\in{\cal L}(Z,X)$ onto a
finite-dimensional subspace $Z_c=X_c\subset X$, such that
$$A\pi_cx=\pi_cAx,\quad\forall x\in X,$$
and such that for
$$Z_h=(I-\pi_c)(Z),\quad X_h=(I-\pi_c)(X),\quad Y_h=(I-\pi_c)(Y),$$
$$A_c=A|_{X_c}\in {\cal L}(X_c),\quad A_h=A|_{X_h}\in {\cal L}(X_h,Z_h),$$
the following statements hold:

\medskip\noindent
(i) there exist $\alpha_+\ge0$ and $\alpha_-\ge0$ such that
$$-\alpha_-\le\hbox{Re}\lambda\le\alpha_+\quad\forall\lambda\in\sigma(A_c);$$

\medskip\noindent
(ii) there exist $\beta_-$ and $\beta_+$, $\beta_\pm>k\alpha_\pm$, such that
for any $\bfe=(\eta_-,\eta_+)$, $\eta_\pm\in[0,\beta_\pm)$, and for
any $f\in BC^{\bfe}(\R,Y_h)$ the linear problem
$$\dot x_h= A_hx_h+f(t),\quad x_h\in BC^{\bfe}(\R,X_h)$$
has a unique solution $x_h=K_hf$, where
$K_h\in{\cal L}(BC^{\bfe}(\R,Y_h),BC^{\bfe}(\R,X_h))$ and
$$\|K_h\|_{\bfe}\le\gamma(\bfe)$$
for a continuous function
$\gamma:[0,\beta_-)\times[0,\beta_+)\to\R_+$.

\medskip
{\it Lemma 8.} Assume ({\bf H}) and $g\in C^0_b(X,Y)$. Let
$\tilde x:\R\to X$ be a solution of (\ref{equa2}), and let
$\bfe=(\eta_-,\eta_+)\in (\alpha_-,\beta_-)\times(\alpha_+,\beta_+)$.
Then the following statements are equivalent:

\medskip\noindent
(i) $\tilde x\in BC^{\bfe}(\R,X)$;

\medskip\noindent
(ii) $\tilde x\in BC^{\bfx}(\R,X),\quad\forall\bfx=(\xi_-,\xi_+),\ \xi_\pm>\alpha_\pm$;

\medskip\noindent
(iii) $\pi_h\tilde x\in C^0_b(R,X_h)$.

\medskip\noindent
The proof is identical to that of Lemma 1 in \cite{VanI92}.

\medskip
{\it Lemma 9.} Assume ($\bf H$) and $g\in C^0_b(X,Y)$. Let
$\tilde x\in BC^{\bfe}(\R,X)$ for some
$\bfe=(\eta_-,\eta_+)\in (\alpha_-,\beta_-)\times(\alpha_+,\beta_+)$.
Then $\tilde x$ is a solution of (\ref{equa2}) if and only if

$$\tilde x(t)=e^{A_ct}\pi_c\tilde x(0)+
\int_0^te^{A_c(t-s)}\pi_cg(\tilde x(s))ds+K_h(\pi_hg(\tilde x))(t),\quad\forall
t\in\R.$$

\noindent
The Lemma is identical to Lemma 2 of \cite{VanI92}.

\medskip
{\it Theorem 4.} Assume ($\bf H$). Then there exist $\delta_0>0$ such
that for all $g\in C^{0,1}_b(X,Y)$, which are globally Lipschitz with the
Lipschitz constant $|g|_{\rm Lip}$ satisfying
\begin{equation}\label{glip}
|g|_{\rm Lip}<\delta_0,
\end{equation}
there exist a unique $\psi\in C^{0,1}_b(X_c,X_h)$ possessing the property that
for all $\tilde x:\R\to X$ the following statements are equivalent:

\medskip\noindent
(i) $\tilde x$ is a solution of (\ref{equa2}) and $\tilde x\in BC^{\bfe}(\R,X)$
for some $\bfe=(\eta_-,\eta_+)\in (\alpha_-,\beta_-)\times(\alpha_+,\beta_+)$;

\medskip\noindent
(ii) $\pi_h\tilde x(t)=\psi(\pi_c\tilde x(t))$ for all $t\in\R$ and
$\pi_c\tilde x:\R\to X_c$ is a solution of the equation
\begin{equation}\label{dxc}
\dot x_c=A_cx_c+\pi_cg(x_c\psi(x_c)).
\end{equation}

\medskip\noindent
As pointed out in \cite{VanI92}, the proof is similar to the
proof of Theorem 1 in \cite{Van89} and is the same as the proof of
Theorem 1 in the present paper.

\medskip
The Theorem implies that, assuming ($\bf H$) and $g\in C^{0,1}_b(X,Y)$
satisfying (\ref{glip}), the problem
$$
\left\{
\begin{array}{l}
\dot x=Ax+g(x)\\
\pi_cx(0)=x_c,\ x\in BC^{\bfe}(\R,X)
\end{array}\right.
$$
with $\bfe=(\eta_-,\eta_+)\in (\alpha_-,\beta_-)\times(\alpha_+,\beta_+)$
has for each $x_c\in X_c$ a unique solution
$$\tilde x(t,x_c)=\tilde x_c(t,x_c)+\psi(\tilde x_c(t,x_c)),$$
where $\tilde x_c(t,x_c)$ is the unique solution of (\ref{dxc}) satisfying
$x_c(0)=x_c$.

\medskip
As in the finite-dimensional case (Section 1), the set
$$M_c=\{x_c+\psi(x_c) | x_x\in X_c\}\subset X$$
is called the {\it global center manifold} of (\ref{equa2}).

\medskip
{\it Theorem 5.} Assume ($\bf H$). Then for any $l\le k$ there exist
$\delta_l>0$, such that if $g\in C^{0,1}_b(X,Y)\cap C^l_b(V_\rho,Y)$,
with $V_\rho=\{x\in X| \|\pi_hx\|<\rho\}$ and $\rho>\|K_h\|_0|\pi_hg|_0$,
\begin{equation}\label{gllip}
|g|_{\rm Lip}<\delta_l
\end{equation}
the mapping $\psi$ given by Theorem 1 belongs to the space $C^l_b(X_c,X_h)$.

\medskip\noindent
Similarly to Theorem 4, the proof follows the proof of Theorem 2 for
finite-dimensional systems.

\medskip
{\it Theorem 6.} Assume ($\bf H$), $g\in C^k(X,Y)$ for $k\ge1$,
$g(0)=0$ and $Dg(0)=0$. Then there exist a neighborhood $\Omega$ of
the origin in $X$ and a mapping $\psi\in C^k_b(X_c,X_h)$ with
$\psi(0)=0$ and $D\psi(0)=0$ such that the following statements hold:

\medskip\noindent
(i) if $\tilde x_c:I\to X_c$ is a solution of (\ref{dxc}) such that
$\tilde x(t)=\tilde x_c(t)+\psi(\tilde x_c(t))\in\Omega$ for all $t\in I$,
then $\tilde x:I\to X$ is a solution of (\ref{equa2});

\medskip\noindent
(ii) if $\tilde x:\R\to X$ is a solution of (\ref{equa2}) such that
$\tilde x(t))\in\Omega$ for all $t\in\R$, then
$$\pi_h\tilde x(t)=\psi(\pi_c\tilde x(t)),\quad\forall t\in\R,$$
and $\pi_c\tilde x:\R\to X_c$ is a solution of (\ref{dxc}).

\medskip\noindent
Unlike in the cases of Theorems 4 and 5, the proof is different from the one for
finite-dimensional systems, since the cut-off function $\chi\in C^k_b(X,\R)$
used in the proof of Theorem 3 does not always exist for a general Banach
space $X$. The proof for infinite-dimensional systems given in \cite{VanI92}
involves construction of a cut-off function from the finite-dimensional
$X_c$ to $\R$.

\bigskip
{\large\bf 3. The Navier-Stokes equation}

\bigskip
Consider the Navier-Stokes equation
\begin{equation}\label{eq_ns}
{\partial{\bf v}\over\partial t}={\bf v}\times(\nabla\times{\bf v})
-\nabla p+\nu\Delta{\bf v}+{\bf f}
\end{equation}
subject to the incompressibility condition
\begin{equation}\label{inc}
\nabla\cdot{\bf v}=0,
\end{equation}
where the force ${\bf f}$ is a smooth bounded function, defined in a bounded
domain $\Omega\subset\R^3$ with a smooth boundary $\partial\Omega$.

\medskip
We assume one of the following boundary conditions:

\medskip\noindent
space-periodic:
\begin{equation}\label{bc1}
{\bf v}({\bf x})={\bf v}({\bf x}+{\bf T}),\quad{\bf T}\in\R^3;
\end{equation}

\medskip\noindent
no-slip:
\begin{equation}\label{bc2}
{\bf v}|_{\partial\Omega}=0.
\end{equation}

\medskip\noindent
Our theory is equally applicable to other commonly used boundary conditions,
e.g. stress-free and periodicity in one (the Taylor-Couette problem)
or two directions (in a layer).

Denote by $\cal F$ the space of functions,
satisfying the boundary conditions (\ref{bc1}) or (\ref{bc2}).

Let ${\bf v}_0$ be a steady solution of
(\ref{eq_ns}), (\ref{inc}) with (\ref{bc1}) or (\ref{bc2}).
For ${\bf v}={\bf v}_0+\bf w$, (\ref{eq_ns}) reduces to
\begin{equation}\label{eq_nsn}
{\partial{\bf w}\over\partial t}=A{\bf w}+N({\bf w}),
\end{equation}
where
$$A{\bf w}=\pi_0(\nu\Delta{\bf w}+{\bf v}_0\times(\nabla\times{\bf w})+
{\bf w}\times(\nabla\times{\bf v}_0)),$$
$$N({\bf w})=\pi_0({\bf w}\times(\nabla\times{\bf w})).$$
We set
$$Z=\{{\bf w}\in{\cal F}\cap(L_2(\Omega))^3|\ \nabla\cdot{\bf w}=0\},$$
denote by $\pi_0$ the orthogonal projection of $(L_2(\Omega))^3$ onto $Z$, and
define
$$X=Z\cap(H_2(\Omega))^3,\quad Y=Z\cap(H_1(\Omega))^3.$$
It is shown in \cite{VanI92} that $A\in{\cal L}(X,Z)$ and $N\in C^\infty(X,Y)$.
Since $A$ is an elliptic operator, for any constant $C$ it has a finite number
of eigenvalues with Re$\lambda>C$ (counting with multiplicities).

Theorem 6 is applicable to the Navier-Stokes equation, if the equation
satisfies the hypothesis ($\bf H$). Decompose the spectrum of the operator $A$,
$\sigma(A)\subset\C$, into a disjoint union of the stable spectrum
$\sigma_s$, the center spectrum $\sigma_c$ and the unstable spectrum
$\sigma_u$, where
$$\sigma_s=\{\lambda\in\sigma\mid\hbox{Re}\lambda<-\beta_-\},$$
\begin{equation}\label{decompf}
\sigma_c=\{\lambda\in\sigma\mid-\alpha_-\le\hbox{Re}\lambda\le\alpha_+\},
\end{equation}
$$\sigma_u=\{\lambda\in\sigma\mid\hbox{Re}\lambda>\beta_+\}$$
with $\beta_\pm>k\alpha_\pm\ge0$. Due to the properties of the operator $A$
stated above, it can be easily examined for a particular bifurcation by
computing several eigenvalues with the largest real parts for the system
linearized in the vicinity of the steady state, whether for a given $k$ such
constants $\alpha_\pm$ and $\beta_\pm$ can be found that (\ref{decompf}) can
be constructed.

If the decomposition (\ref{decompf}) can be constructed, the Navier-Stokes
equation satisfies the following hypothesis $(\Sigma)$:

\medskip\noindent
There exist $\alpha'_\pm\ge0$ and $\beta'_\pm>k\alpha'_\pm$ such that

\medskip\noindent
(i) $\sigma(A)\cap[\alpha'_-,\alpha'_+]\times i\R$ consists of a finite number of
isolated eigenvalues, each associated with a finite-dimensional generalized eigenspace;

\medskip\noindent
(ii) $([\beta'_-,\alpha'_-]\cup[\alpha'_+,\beta'_+])\times i\R\subset\rho(A)$;

\medskip\noindent
(iii) there exist constants $\omega_0>0$, $C>0$ and $\alpha\in[0,1)$ such
that for all $\omega\in\R$ with $|\omega|\ge\omega_0$ we have
$i\omega\in\rho(A)$,
$$\|(i\omega-A)^{-1}\|_{{\cal L}(Z)}\le{C\over|\omega|}\hbox{ and }
\|(i\omega-A)^{-1}\|_{{\cal L}(X,Y)}\le{C\over|\omega|^{1-\alpha}},$$
where $\rho(A)$ is the resolvent set of $A$.

\medskip\noindent
In \cite{VanI92} the hypothesis $(\Sigma)$ with $\alpha'_\pm=0$ is employed,
and it is shown that it holds for the Navier-Stokes equation.
Their proof can be easily extended for our case (the condition (\ref{decompf})
is required to allow for non-vanishing $\alpha'_\pm$). A trivial modification
of arguments of \cite{VanI92} proves $(\Sigma)\Rightarrow(\bf H$). Thus
Theorem 6 is applicable for the Navier-Stokes equation under the condition
(\ref{decompf}).

The equation (\ref{eq_ns}) involves the parameter $\nu$ (and possibly
others, e.g. included into the force $\bf f$). Denote by $\mu$ all
parameters of the system. CM can be made parameters dependent by the standard
\cite{GucHol} extension of the system by considering the parameters as
variables and setting $\dot\mu=0$. Evidently, Theorem 6 is applicable to the
extended system, if it is applicable to the original one.

\pagebreak
{\large\bf Conclusion}

\bigskip
We have proved CM theorems, including the one for infinite-dimensional systems,
under less restrictive assumptions than those required by existing theorems.
Although the proof is just a modification of the existing proofs
\cite{Van89,VanI92}, the new variant of the theorem (Theorem 6) is important
for applications, providing a more powerful tool for investigation of
bifurcations in dynamical systems of infinite dimensions. Its advantage was
demonstrated by applying our theorem to the ABC-forced Navier-Stokes equation
\cite{PoAsHa,pod05}.

The demonstration that the theorem is applicable for the Navier-Stokes equation
(if additional inequalities for eigenvalues of the linearization in a vicinity
of a steady state hold) relies only on the fact that the linearization
is an elliptic operator. Thus, it can be easily extended to accommodate other
boundary conditions, the Rayleigh-B\'enard convection, magnetohydrodynamic and
other systems.

\bigskip
{\bf Acknowledgments}

\bigskip
I am most grateful to Professor I.Labouriau for her stimulating remarks.
This work has been partly financed by the grant from the Russian
Foundation for Basic Research 04-05-64699.


\begin{thebibliography}{99}

\bibitem{arm}
Armbruster,~D., Guckenheimer,~J. \& Holmes,~P. 1988
Heteroclinic cycles and modulated travelling waves in systems
with $O(2)$ symmetry. {\em Physica D} {\bf 29}, 257--282.

\bibitem{AshPod}
Ashwin,~P. \& Podvigina,~O. 2003
Hopf bifurcation with cubic symmetry and instability of ABC flow.
{\em Proc. Royal Soc. London A} {\bf 459}, 1801-1827.

\bibitem{ChosIos}
Chossat,~P. \& Iooss,~G. 1994
{\em The Couette--Taylor Problem}.
Appl. Math. Sci. {\bf 102}, Springer-Verlag, New York.

\bibitem{Hen}
Henry,~D. 1981
{\em Geometrical theory of semilinear parabolic equations}.
Lecture Notes in Mathematics {\bf 840}, Springer-Verlag, New York.

\bibitem{Ioo}
Iooss,~G. 1979
{\em Bifurcation of Maps and Applications}.
North-Holland Math. Studies {\bf 36}, North-Holland Publishing Company

\bibitem{Gol85}
Golubitsky,~M., \& Schaeffer,~D. 1985
{\em Singularities and Groups in Bifurcation Theory. Volume 1}.
Appl. Math. Sci. {\bf 51}, Springer-Verlag, New York.

\bibitem{Gol88}
Golubitsky,~M., Stewart,~I.N., \& Schaeffer,~D. 1988
{\em Singularities and Groups in Bifurcation Theory. Volume 2}.
Appl. Math. Sci. {\bf 69}, Springer-Verlag, New York.

\bibitem{GucHol}
Guckenheimer,~J., \& Holmes,~P. 1993
{\em Nonlinear Oscillations, Dynamical Systems and Bifurcations
of Vector Fields}. Appl. Math. Sci. {\bf 42}, Springer-Verlag, New York.

\bibitem{MarMc}
J. Marsden,~J., \& McCracken,~M. 1976
{\em The Hopf Bifurcation and its Applications}. Springer-Verlag, New York.

\bibitem{pod94}
Podvigina,~O., \& Pouquet,~A. 1994
On the non-linear stability of the 1:1:1 ABC flow.
{\em Physica D} {\bf 75}, 471--508.

\bibitem{pod99a}
Podvigina,~O.M. 1999
Spatially-periodic steady solutions to the three-dimensional Navier-Stokes
equation with the ABC-force. {\em Physica D} {\bf 128}, 250--272.

\bibitem{PoAsHa}
Podvigina,~O., Ashwin,~P. \& Hawker,~D.  2005
Modelling instability of ABC flow using a mode interaction between steady and
Hopf bifurcations with rotational symmetries of the cube, submitted to
{\em Physica D}.

\bibitem{pod05}
O.M.~Podvigina 2005
Investigation of the ABC flow instability with application of center manifold
reduction, accepted in {\em Dynamical Systems}.

\bibitem{projo}
Proctor,~M.R.E. \& Jones,~C.A. 1988
The interaction of two spatially resonant patterns in
thermal convection. Part 1. Exact 1:2 resonance.
{\em J. Fluid Mech.} {\bf 188}, 301--335.

\bibitem{Shu}
Shub,~M. 1987
{\em Global Stability of Dynamical Systems}. Springer-Verlag, New York.

\bibitem{Van89}
Vanderbauwhede,~A. 1989
Centre Manifolds, Normal Forms and Elementary Bifurcations.
{\em Dynamics Reported. Volume 2}. Eds. U. Kirchgraber and H.O. Walther.
John Willey \& Sons.

\bibitem{VanI92}
Vanderbauwhede,~A. \$ Iooss,~G. 1992
Centre Manifold Theory in Infinite Dimensions.
{\em Dynamics Reported: expositions in dynamical systems. Volume 1}.
Eds. C.K.R.T.Jones, U. Kirchgraber and H.O. Walther. Springer-Verlag, Berlin.
\end{thebibliography}
\end{document}